\documentclass{article}

\PassOptionsToPackage{numbers, compress}{natbib}
\usepackage[main,final]{neurips_2026}

\usepackage{listings}
\usepackage[utf8]{inputenc}
\usepackage[T1]{fontenc}
\usepackage{hyperref}
\usepackage{url}
\usepackage{booktabs}
\usepackage{amsmath,amssymb,amsfonts}
\usepackage{nicefrac}
\usepackage{microtype}
\usepackage{xcolor}
\usepackage{multirow}
\usepackage{graphicx}
\usepackage{algorithm}
\usepackage{algpseudocode}
\usepackage{caption}
\usepackage{subcaption}
\usepackage[shortlabels]{enumitem}
\usepackage{tabularx}
\usepackage{array}
\usepackage{xspace}
\usepackage{caption}
\usepackage{tikz}
\usepackage{xcolor}

\usepackage{booktabs}
\usepackage{wrapfig}

\usepackage{booktabs}
\usepackage{needspace}
\newcommand{\splitzip}{SplitZip\xspace}

\title{SplitZip: Ultra Fast Lossless KV Compression for Disaggregated LLM Serving}

\author{%
  Yipin Guo, \quad Siddharth Joshi \\
  University of Notre Dame \\
  IN, USA \\
  yguo23@nd.edu, \quad sjoshi2@nd.edu \\
}

\begin{document}
\maketitle
\begin{abstract}
Contemporary systems serving large language models (LLMs) have adopted prefill-decode disaggregation to better load-balance between the compute-bound prefill phase and the memory-bound decode phase. Under this design, prefill workers generate a KV cache that must be transferred to decode workers before token generation can begin. With these workers residing on different physical systems, this transfer becomes a significant bottleneck to serving LLMs at scale. This bottleneck gets exacerbated for long-input and agentic workloads.  Existing lossless codecs are not suited to this setting as they primarily target offline weight compression, run on the CPU, or use variable-length coding whose decompression is fast but compression is too slow to keep up with KV production during prefill. We introduce SplitZip, a GPU-friendly lossless compressor for KV cache transfer that preserves KV tensors bitwise and integrates into existing serving frameworks without changes to model execution.  SplitZip exploits redundancy in floating-point exponents of KV activations, encoding the most frequent exponent values with fixed-length codes and routing rare exponents through a sparse escape stream of (position, value). 
An offline calibrated top-16 exponent codebook eliminates online-histogramming, while the regular dense path and sparse escape correction make both encoding and decoding efficient on GPUs. On real BF16 activation tensors, SplitZip achieves 613.3 GB/s compression throughput and 2181.8 GB/s decompression throughput, substantially outperforming prior lossless compressors on the latency-critical codec path. End-to-end transfer experiments show up to 1.32× speedup for BF16 KV cache transfer, 1.30× speedup for TTFT, and 1.23× increase on Request Throughput.
The same approach extends to FP8 KV caches, providing up to 1.14× compression over native E5M2.
Code is available at \url{https://github.com/Intelligent-Microsystems-Lab/SplitZip}.

\end{abstract}

\section{Introduction}
\label{sec:intro}
Prefill-decode (PD) disaggregation has become a widely adopted systems design pattern~\cite{zhong2024distserve,jia2025disaggregated} for large-scale LLM serving because it enables specialization during two fundamentally different phases of inference: prefill is primarily compute-intensive, whereas decode is primarily memory-bandwidth-intensive. This system direction has influenced the larger serving ecosystem, having been widely adopted by open frameworks such as vLLM~\cite{vllm2025}, SGLang~\cite{sglang2025}, and Dynamo~\cite{dynamo2025}. However, although PD disaggregation improves resource utilization and scheduling flexibility, its efficiency depends on KV Cache transfer between the prefill and decode stages. When prefill and decode are deployed on different nodes or different clusters, the KV Cache generated during prefill must be transmitted quickly enough to keep decode continuously fed. In tightly coupled deployments, this transfer can often be sustained by a high-bandwidth RDMA fabric. In more realistic production settings, however, prefill and decode are frequently separated across deployment boundaries for reasons such as resource isolation, cluster management, elasticity, distinct accelerator architectures, or independent scaling. In such settings, KV Cache transfer must traverse slower and more constrained inter-node or inter-cluster links, making communication overhead a growing bottleneck~\cite{qin2026prefillasaservice}.

This transfer bottleneck is especially pronounced for long-input workloads, such as document-level question answering, codebase understanding, and multi-document summarization, where prefill produces a large KV cache that must be transferred to the decode workers. Recent serving deployments increasingly separate prefill and decode across different hardware tiers or even different datacenters, where inter-cluster bandwidth (typically 50-100 GB/s) is an order of magnitude lower than the intra-rack RDMA fabrics (400+ GB/s within NVLink) that earlier disaggregation work assumed~\cite{qin2024mooncake}. Therefore, accelerating KV Cache transfer is a key requirement for making PD disaggregation efficient in standard large-scale serving deployments.

To mitigate this, approaches have explored lossy KV cache compression, such as pruning\cite{zhang2023h2o,li2024snapkv}, quantizing\cite{hooper2024kvquant,zhang2024kv1bit}, or selectively retaining KV states\cite{beltagy2020longformer,xiao2024streamllm}, as well as system-level solutions that depend on specialized hardware assumptions\cite{fan2026zipserv} or changes to the serving paradigm\cite{qin2024mooncake}. Although lossy methods can achieve higher compression ratios, they alter model outputs in ways that require per-deployment validation and accumulate errors across stored KV tokens, an effect that worsens at the long contexts and multi-step reasoning we target\cite{kvlinc,pmkvq}. We propose a lossless compression approach for KV Cache transfer in conventional PD-disaggregated deployments called \textbf{\splitzip}. Our goal is to reduce communication volume while preserving exact model semantics, thereby improving transfer efficiency without sacrificing model ability or changing any tensor values. 

\begin{figure}[t]
    \centering
    \includegraphics[width=\textwidth]{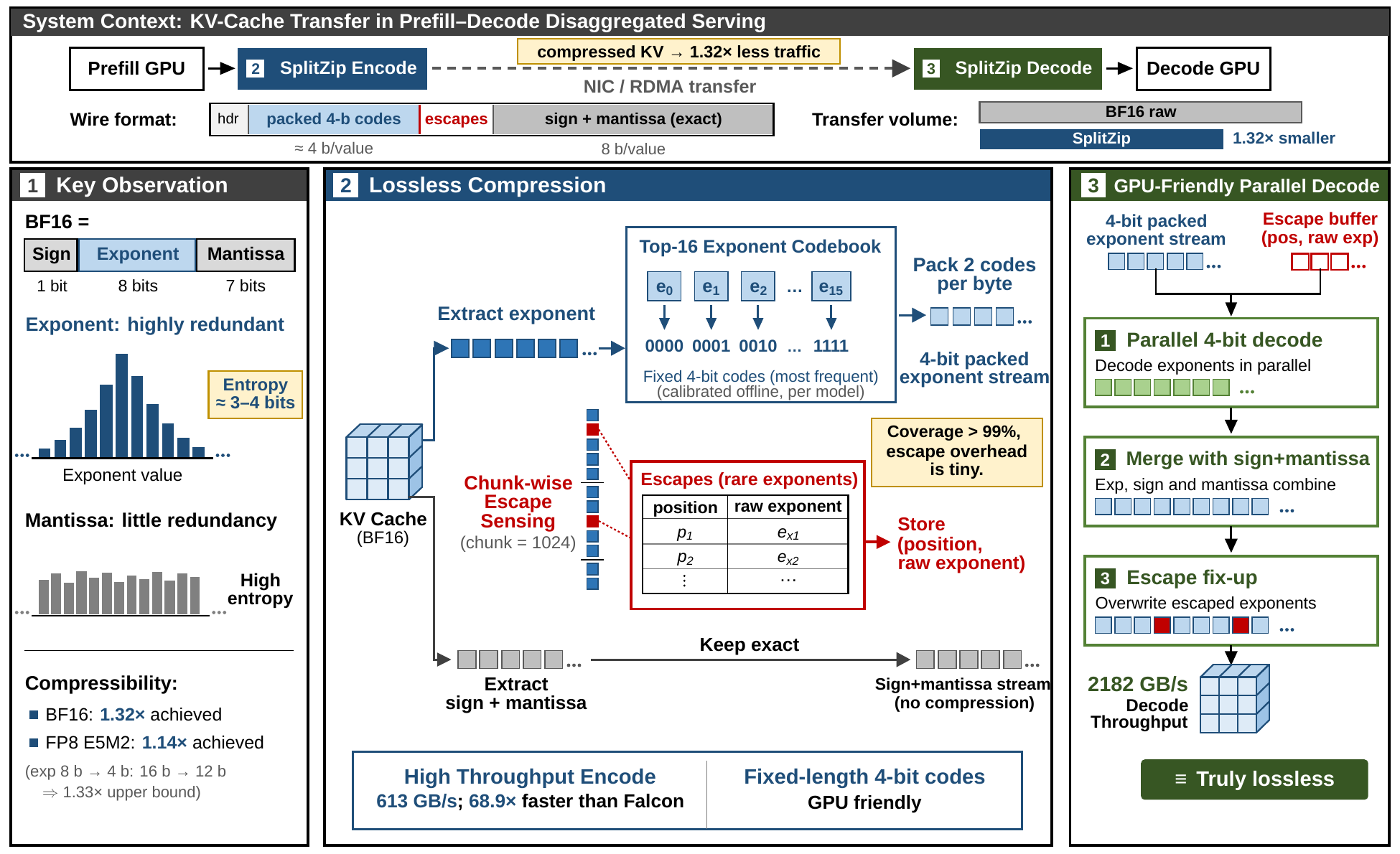}
    \caption{\textbf{Overview of SplitZip for lossless KV cache compression.} SplitZip exploits the redundancy of the BF16 exponent field while keeping sign and mantissa exact. It encodes the most frequent exponent values with fixed 4-bit codes and stores rare values in a small escape buffer with their positions and raw exponents. This design enables GPU-friendly parallel decoding and reconstructs the original KV cache bit-exactly.}
    \label{fig:main}
    \vspace{-0.5em}
\end{figure}

Prior work has shown that, in BF16 model weights, redundancy is concentrated in the exponent field, while the mantissa has limited compressibility~\cite{dfloat11}. We find that this also holds for activations, including KV tensors in PD-disaggregated serving. Across common LLMs, we find the exponent entropy to be typically 3--4 bits, indicating roughly 4 bits of headroom for lossless compression (see Fig.~\ref{fig:main} left).

Previous efforts have used Huffman coding\cite{yubeaton2025huffllm,dfloat11,agrawal202sshuffman} to losslessly compress weight tensors. Huffman codes are entropy-optimal in the limit, but their variable-length bitstreams produce irregular memory access patterns that map poorly to GPU execution. Blockwise bitstream processing\cite{yubeaton2025huffllm} and variable-length coding\cite{agrawal2026quadlength} reduce this overhead, but variable codes still impose sequential dependencies during decoding that limit parallel throughput. Our approach, shown in Fig.~\ref{fig:main}, uses fixed 4-bit codes for the most frequent exponent values, and stores the rare uncovered cases as escapes with their positions and original values. In practice, 4-bit coding covers about 99\% of cases, as shown in Table~\ref{tab:exponent}, so escape overhead is very small. At the same time, the fixed-length design maps naturally to GPU parallelism, enabling much higher encoding and decoding throughput while remaining fully lossless. 

Our contributions can be summarized as:

\begin{itemize}
    \item \textbf{Characterizing redundancy in KV activations}. We show that the BF16 exponent redundancy previously documented in model weights also appears in KV activations, and that it is consistent across six model spanning dense, MoE, and hybrid architectures, across multiple calibration datasets, and across nearly all transformer layers we tested. Existing approaches achieve high compression ratios, but operate at sub-GBps rates, too slow to keep up with prefill generation~\cite{dfloat11,zipnn}.
 
    \item \textbf{\splitzip for online encode-decode}. We present a fixed-length codec that achieves 613.3 GB/s encoding and 2181.8 GB/s decoding throughput on real BF16 KV activations, the highest measured rates among lossless floating-point compressors we evaluated. Our ablations show that explicit escape positions, despite a small metadata overhead, deliver 3.5$\times$ higher decode throughput owing to divergence-free GPU execution.

    \item \textbf{End-to-end validation in a production serving stack}. We integrate \splitzip into SGLang~\cite{sglang2025} on top of Mooncake~\cite{qin2024mooncake} transport. This delivers up to 1.32$\times$ faster BF16 KV cache transfer, 1.30$\times$ faster TTFT, and 1.23$\times$ higher request throughput. The same approach delivers 1.14$\times$ additional compression over native E5M2 FP8 KV Caches.
\end{itemize}

\section{Related Work}
\label{sec:related}

\textbf{Disaggregated PD serving.}
DistServe~\cite{zhong2024distserve} was one of the first to separate prefill and decode to improve goodput under latency constraints, while Splitwise~\cite{splitwise} pipelines phase-split inference across heterogeneous GPUs.
Mooncake~\cite{qin2024mooncake} introduces a KV cache-centric disaggregated architecture with a Transfer Engine for RDMA and TCP transport.
FlowKV~\cite{flowkv} focuses on fragmented KV memory transfer and segmented kernel launches.
These systems reduce scheduling or transfer overhead, but they do not employ compression for KV transfer itself.

\textbf{Lossless LLM compression.}
General GPU compression libraries such as nvCOMP\cite{nvidia2026nvcomp} provide LZ4, Cascaded, and entropy-coding primitives.
They are valuable for generic data, but floating-point KV tensors can benefit from format-aware codes: although the BF16 mantissa stream is hard to compress, the exponent stream has structure.
DFloat11~\cite{dfloat11} compresses BF16 model weights by coding only the exponent bits, improving decode throughput. 
ZipNN~\cite{zipnn} studies lossless compression for weights, checkpoints, and low-precision tensors. 
ZipLLM~\cite{zipllm} combines tensor-level deduplication with lossless compression for efficient LLM storage.
They all employ compression methods based on Huffman coding. 
More recently, Falcon\cite{Li2025falcon} and ZipServ\cite{fan2026zipserv} utilize multi-level fixed-length encoding and  execution to the GPU, thereby further enhancing throughput.

\splitzip adopts fixed-length coding from this line of work but targets a different regime. The latency-critical KV transfer path in disaggregated serving requires both compression and decompression throughput to be optimized, leading to different performance tradeoffs.


\begin{table}[t]
\centering
\small
\caption{BF16 KV value exponent statistics across different model families and architectures. Top-16 coverage exceeds 99\% for high-frequency exponents, motivating our 4-bit exponent coding.}
\vspace{0.2em}
\label{tab:exponent}
\begin{tabular}{llcccc}
\toprule
Model & Family & Top-8 & Top-16 & Entropy & Realized SplitZip CR\\
\midrule
Qwen3-30B-A3B & Qwen-MoE & 83.5\% & 99.3\% & 3.59\,b & $1.31\times$\\
Qwen3-32B & Qwen & 90.1\% & 99.8\% & 3.29\,b & $1.32\times$\\
Qwen3-Next-80B-A3B & Qwen-Hyper & 96.0\% & 99.9\% & 2.89\,b & $1.33\times$\\
Llama-3.1-70B & Llama & 89.7\% & 99.5\% & 3.41\,b& $1.32\times$\\
Llama-3-8B & Llama & 90.9\% & 99.6\% & 3.30\,b  & $1.32\times$\\
Phi-2 & Phi & 92.3\% & 99.6\% & 3.11\,b &  $1.32\times$\\
\bottomrule
\vspace{-2.5em}
\end{tabular}
\end{table}

\section{\splitzip}
\label{sec:method}

\subsection{KV Exponent Entropy}
\label{sec:method_motivation}

Table~\ref{tab:exponent} characterizes the exponent distribution of KV cache across multiple model families. Despite differences in architecture and numerical range, we consistently observe low exponent entropy, ranging from 2.89 to 3.59 bits vs. the 8-bit BF16 exponent. The distribution is also highly concentrated, with the most frequent 16 exponent values covering  $\ge$99.3\% of all exponent entries across all evaluated models, including dense, MoE, and hybrid architectures. Top-8 coverage varies from 83.5\% on Qwen3-30B-A3B to 96.0\% on Qwen3-Next-80B-A3B, indicating that a 3-bit codebook is less robust across activation distributions.
SplitZip therefore uses a Top-16 design (see Sec.~\ref{sec:method_top16} and Sec.~\ref{sec:ablation_topk} for details). 

\subsection{Method}
\label{sec:method_overview}

SplitZip combines fixed-length exponent coding with explicit escape capture, as illustrated in Figure~\ref{fig:main}. For a BF16 value represented as a 16-bit integer $x_i$, SplitZip extracts

\[
    e_i = (x_i \gg \texttt{7}) \mathbin{\&} \texttt{0xff},
    \qquad
    a_i = ((x_i \gg \texttt{8}) \mathbin{\&} \texttt{0x80}) \;\mid\; (x_i \mathbin{\&} \texttt{0x7f}),
\]
where $e_i$ is the 8-bit exponent and $a_i$ is the exact sign--mantissa byte.
Given a decoded exponent $\hat e_i$, the original BF16 bit pattern is reconstructed as
\[
    \hat x_i =
    ((a_i \mathbin{\&} \texttt{0x80}) \ll \texttt{8})
    \;|\;
    (\hat e_i \ll \texttt{7})
    \;|\;
    (a_i \mathbin{\&} \texttt{0x7f}).
\]

The sign and mantissa bits are stored losslessly in an 8-bit stream. SplitZip uses a calibrated codebook for the exponent stream, containing the 16 most frequent exponent values.  Each common exponent is encoded using a 4-bit code, with two 4-bit codes packed into a byte. Exponent values outside the top-16 are treated as escape values. 
For each escape value, SplitZip records its chunk-relative element position as a 16-bit unsigned integer and its raw exponent value as an 8-bit integer.

Let $\mathcal{C}=\{c_0,\ldots,c_{15}\}$ be the calibrated top-16 exponent set and $\mathcal{E} = \{ i \mid e_i \notin \mathcal{C} \}$ be the escape set. 
For a tensor with $N$ BF16 elements and $M=|\mathcal{E}|$ escaped exponents, SplitZip stores
\[
    B_{\mathrm{SZ}}
    =
    \underbrace{N}_{\text{sign--mantissa}}
    +
    \underbrace{\frac{N}{2}}_{\text{4-bit exponent codes}}
    +
    \underbrace{3M}_{\text{escape positions and values}}
    =
    N\left(\frac{3}{2}+3\epsilon\right)
    \qquad \mathrm{Bytes}.
\]
Here the $\epsilon=M/N$ is the escape rate. 
Since the uncompressed BF16 tensor uses $B_{\mathrm{raw}}=2N$ bytes, the compression ratio is
\[
    \rho
    =
    \frac{B_{\mathrm{raw}}}{B_{\mathrm{SZ}}}
    =
    \frac{2}{\frac{3}{2}+3\epsilon}.
\]
As escape rates decrease, $\rho$ approaches $4/3$ asymptotically.

The first two terms form the compression path, while the final term is the correction stream. 
When the top-16 exponent coverage is high (covers greater than 99\%), $M$ is very small, so the escape overhead is negligible compared with the reduction from replacing each 8-bit exponent with a 4-bit code.

The encoding path consists of two passes.
The first pass performs dense encoding by extracting exponents, mapping them to 4-bit codes, and packing the exponent and sign--mantissa streams.
Escape counting is fused into this pass using a marked lookup table, avoiding an additional scan.
After a prefix sum determines the compact output offsets from per-chunk escape counts, the second pass collects escape positions and their original exponent values.
This design avoids both worst-case buffer over-provisioning and costly atomic allocation, as evaluated in Sec.~\ref{sec:two_pass}.

To accelerate dense encoding, we introduce \textit{Quad64 vectorized encoding}, which processes four BF16 values within a single 64-bit operation.
This optimization reduces encoding overhead and improves throughput by $1.5\times$ over the unfused baseline.

Decoding follows the reverse order. 
SplitZip first unpacks each byte into two 4-bit codes, maps each code through a 16-entry decoding lookup table, and reconstructs a BF16 bit pattern by combining the decoded exponent with the stored sign--mantissa bits. 
Escaped elements may initially decode to an incorrect common exponent; SplitZip then overwrites them through the sparse correction stream: for each recorded escape position, it overwrites the reconstructed exponent with the exact raw exponent value for lossless reconstruction. 


\subsection{Calibration}
\label{sec:method_calibration}

Computing an exact exponent histogram during every compression call is expensive, instead, SplitZip performs a one-time calibration step on a small representative dataset. 
During calibration, SplitZip extracts all exponent values from the calibration tensors, counts their frequencies, selects the top-16 exponents, and constructs three lookup tables: an encoding table from exponent values to 4-bit codes, a decoding table from 4-bit codes back to exponent values, and a membership table for detecting escapes. This removes histogram construction from the online compression path. 

Although the codebook is selected using a calibration dataset, our experiments show that the selected high-frequency exponent set generalizes well across inputs. 

\subsection{Top-16 Instead of Top-15}
\label{sec:method_top16}

A natural alternative for identifying escape values is to reserve one codeword as an explicit escape token.
This top-15-plus-escape-token design allows the decoder to infer escape positions directly from the packed exponent-code stream, thereby reducing the escape positions overhead.  However, this design also reduces the number of directly represented common exponents from 16 to 15 and introduces special-case handling into the dense decode path.

SplitZip instead uses all 16 possible 4-bit codes for common exponent values. 
Escaped exponents are assigned a dummy code in the packed stream to be overwritten using the explicit escape-position array. 
It increases codebook coverage by retaining the 16th most frequent exponent and preserves a uniform dense decoding path: every element is decoded by the same 4-bit lookup operation. 




\subsection{Quad64 Vectorized Encoding}

Processing scalar BF16 values incurs high instruction and memory overhead. We migitate this through a vectorized encoding. SplitZip uses a Quad64 vectorized encoding kernel, where, 
instead of processing BF16 values as individual elements or pairs, Quad64 reinterprets every four BF16
values as one 64-bit word.
Each program loads one packed 64-bit value, extracts four 16-bit BF16 words, maps their four exponents through the 4-bit codebook, packs the four 4-bit codes into two bytes, and simultaneously packs the four sign--mantissa bytes into one 32-bit word.
This reduces the number of memory instructions and makes the dense output streams more naturally
aligned for GPU execution.

For the default chunk size of 1024, Quad64 is further fused with escape counting.
The optimized kernel not only writes the packed exponent stream and sign--mantissa stream, but also
uses a marked lookup table to detect uncommon exponents and accumulate per-chunk escape counts.
This avoids an additional full pass over the input for escape counting.
\section{Experiments}
\label{sec:experiments}

\subsection{Setup}
    \label{sec:setup}

    \textbf{Platform and timing protocol.}
    Experiments are conducted on server with NVIDIA H200 GPU and dual-socket Intel Xeon Platinum 8468 CPU. The server includes Mellanox ConnectX-7 MT2910 InfiniBand/RDMA adapters for high-speed KV transfer, as well as Mellanox ConnectX-4 Lx and Intel X550 Ethernet adapters. We report compression and decompression throughput in GB/s, where the byte count is the uncompressed BF16 tensor size. Each codec benchmark first verifies bitwise round-trip correctness and then times repeated executions after warm-up. Codec throughput is measured over 10 runs, and tables report the average with measured fluctuation when available. Mooncake\cite{qin2024mooncake} transfer experiments report measured KV-transfer time; the Qwen3-32B\cite{yang2025qwen3} breakdown uses additive accounting for encode, compressed transfer, and decode under the RoCE 4$\times$200G configuration with 700~Gb/s effective payload bandwidth.
    
    \textbf{Datasets and workloads.}
    All main experiments use authentic BF16 KV cache activations. For codec and ablation measurements, we use Qwen3-32B KV tensors assembled into a 256M BF16 workload on wikitext2-test and use chunked escape-value capture with chunk size 1024 (refer to Appendix \ref{sec:ablation_chunksize} for ablation). For Mooncake transfer, we evaluate Llama-3-8B\cite{grattafiori2024llama3} and Qwen3-30B-A3B under four sweeps: fixed batchSize 1 with sequence length from 512 to 131K, fixed batchSize 16 with sequence length from 128 to 64K, fixed sequence length 1024 with batchSize from 1 to 256, and fixed sequence length 32768 with batchSize from 1 to 128. For the Qwen3-32B transmission breakdown, we report sequence lengths 2048, 16384, and 65536. We also instrument SGLang on Qwen3-32B with the same sweep structure to record TTFT and serving throughput. Unless stated otherwise, the BF16 Top-16 codebook is calibrated on the WikiText-2\cite{merity2016wikitext2} training subset.
    
    \textbf{Baselines and ablations.}
    For BF16 codec comparison, we evaluate nvCOMP LZ4, nvCOMP Cascaded, nvCOMP Bitcomp\cite{nvidia2026nvcomp}, DFloat11\cite{dfloat11}, ZipNN\cite{zipnn}, ZipServ\cite{fan2026zipserv}, Falcon\cite{Li2025falcon}, and \splitzip. For transfer-time studies, the native baseline moves raw BF16 KV cache bytes, while \splitzip sends the compressed payload and accounts for encode and decode where applicable. The ablation suite covers BF16 Top-8 3-bit coding versus Top-16 4-bit coding, explicit escape-position metadata versus sentinel-only escape discovery, pre-calibrated versus dynamic Top-16 codebooks, per-tensor/per-token/per-channel calibration granularity, chunk-size sensitivity, codec-stage breakdown, and codebook stability across layers and between K and V caches. Unless otherwise specified, ablations use Qwen3-32B BF16 KV values.

\subsection{Results}
\label{sec:exp_results}\vspace{-0.5em}
\begin{table*}[t]
\centering
\small
\caption{Comparison of encoding/decoding throughput and compression ratio of various compression methods. SplitZip delivers high codec throughput while maintaining a respectable compression ratio.}
\label{tab:baseline_comparison}
\setlength{\tabcolsep}{4pt}
\renewcommand{\arraystretch}{1.08}
\begin{tabular}{l|c|cc}
\toprule
Method & Ratio & Encode(GB/s) & Decode(GB/s) \\
\midrule
nvCOMP LZ4\cite{nvidia2026nvcomp} & 1.019 & $13.4 \pm 0.4$ & $137.1 \pm 8.8$  \\
nvCOMP Cascaded\cite{nvidia2026nvcomp} & 1.000 & $111.8 \pm 0.8$ & $155.2 \pm 5.6$ \\
nvCOMP Bitcomp\cite{nvidia2026nvcomp} & 1.056 & $341.5 \pm 7.1$ & $147.7 \pm 4.8$  \\
ZipNN\cite{zipnn} & 1.515 & 1.2 & 1.7  \\
DFloat11\cite{dfloat11} & 1.423 & $4.0e\!-\!03 \pm 5.0e\!-\!05$ & $468.2 \pm 2.5$ \\
Falcon\cite{Li2025falcon} & 1.428 & $8.9 \pm 1.8$ & $14.4 \pm 3.3$ \\
ZipServ - CPP \cite{fan2026zipserv} & 1.236 & $0.1 \pm 2.1e\!-\!04$ & $499.5 \pm 0.0$ \\
ZipServ - Kernel \cite{fan2026zipserv} & 1.236 & N/A & $1260.9 \pm 1.4$  \\
\textbf{SplitZip} & 1.324 & \textbf{$613.3 \pm 2.6$} & \textbf{$2181.8 \pm 38.5$}\\
\bottomrule
\end{tabular}
\vspace{-1.0em}
\end{table*}
    \begin{figure}[t]
        \centering
        \includegraphics[width=\textwidth]{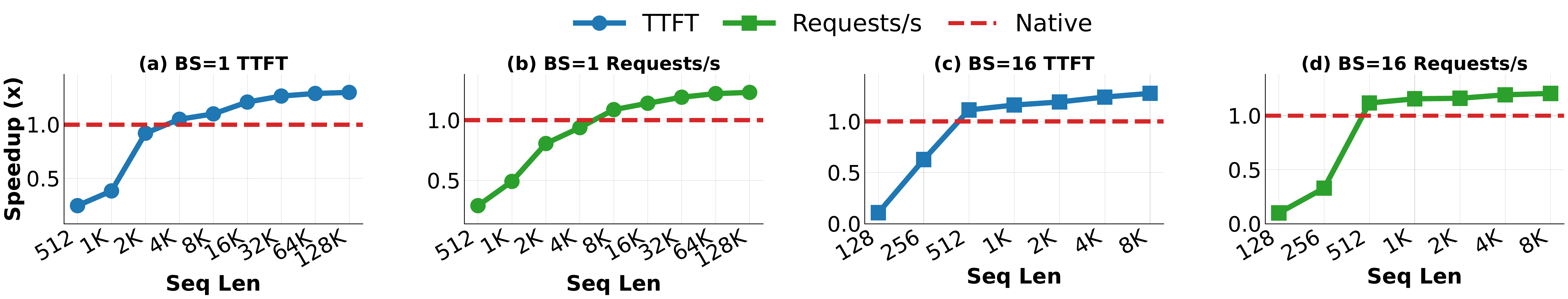}
        \caption{
        End-to-end speedup on Qwen3-32B across sequence-lengths with SplitZip in SGLang~\cite{sglang2025}.}
        \label{fig:sglang_speedup}
        \vspace{-1.5em}
    \end{figure}

    \paragraph{Codec-level performance.}
    Table~\ref{tab:baseline_comparison} compares SplitZip with existing lossless compressors.
    The table includes GPU-oriented baselines evaluated during the revision, including nvCOMP LZ4, Cascaded, and Bitcomp, as well as Falcon and ZipServ.
    For ZipNN, we report the throughput numbers from its original paper.
    For ZipServ, the CPP row corresponds to its CPU computation kernel, while Kernel corresponds to its GPU computation kernel.
    
    SplitZip reaches $613.3 \pm 2.6$~GB/s encoding and $2181.8 \pm 38.5$~GB/s decoding, higher than all measured baselines on both encode and decode.
    Specifically, SplitZip improves encode throughput by $1.8\times$ over nvCOMP Bitcomp and improves decode throughput by $14.1\times$ over nvCOMP Cascaded, the fastest nvCOMP encoder and decoder respectively.
    Compared with ZipServ, SplitZip removes the CPU-side compression bottleneck and achieves $1.7\times$ higher decode throughput than the GPU kernel path.
    Falcon is also GPU-optimized, yet SplitZip achieves $68.9\times$ higher encoding throughput.
    Meanwhile, SplitZip still maintains a strong compression ratio: it is $7.1\%$ higher than ZipServ when including metadata overhead, and only $7.0\%$ lower than DFloat11, a Huffman-based method. 
    
    We also study performance across multiple GPU generations (details in Appendix \ref{sec:more_gpu}), with SplitZip consistently maintaining the highest encoding throughput, improving over the strongest available encoder by 1.8$\times$ to 3.0$\times$ depending on the platform. Decoding behavior is more hardware-dependent: SplitZip leads on every platform, but the margin grows on newer GPUs with higher memory bandwidth.
    This trend highlights the hardware-friendly nature of \splitzip: its decoding throughput is largely governed by the available GPU memory bandwidth, while the compute portion is sufficiently optimized and does not become the bottleneck.


    SplitZip can also be extended to FP8 KV caches by adapting the exponent coding scheme to the FP8 format. For E5M2, the 5-bit exponent field can be compressed using 4-bit Top-16 code under the same fixed-code-plus-escape framework as BF16. Although FP8 already halves the KV cache size compared with BF16, leaving less room for additional lossless compression, E5M2 still exposes meaningful exponent redundancy. In our Qwen3-32B workload, E5M2 achieves $1.14\times$ with a $0.16\%$ escape rate. This is approximately 88\% of the theoretical maximum compression for E5M2 (from the 6.19-bit raw byte entropy). The lower escape rate substantially improves codec throughput: Top-16 reaches $249.7$~GB/s encoding and $564.9$~GB/s decoding throughput. See Appendix \ref{sec:fp8_result} for details.
    
    \begin{figure}[t]
        \centering
        \includegraphics[width=\textwidth]{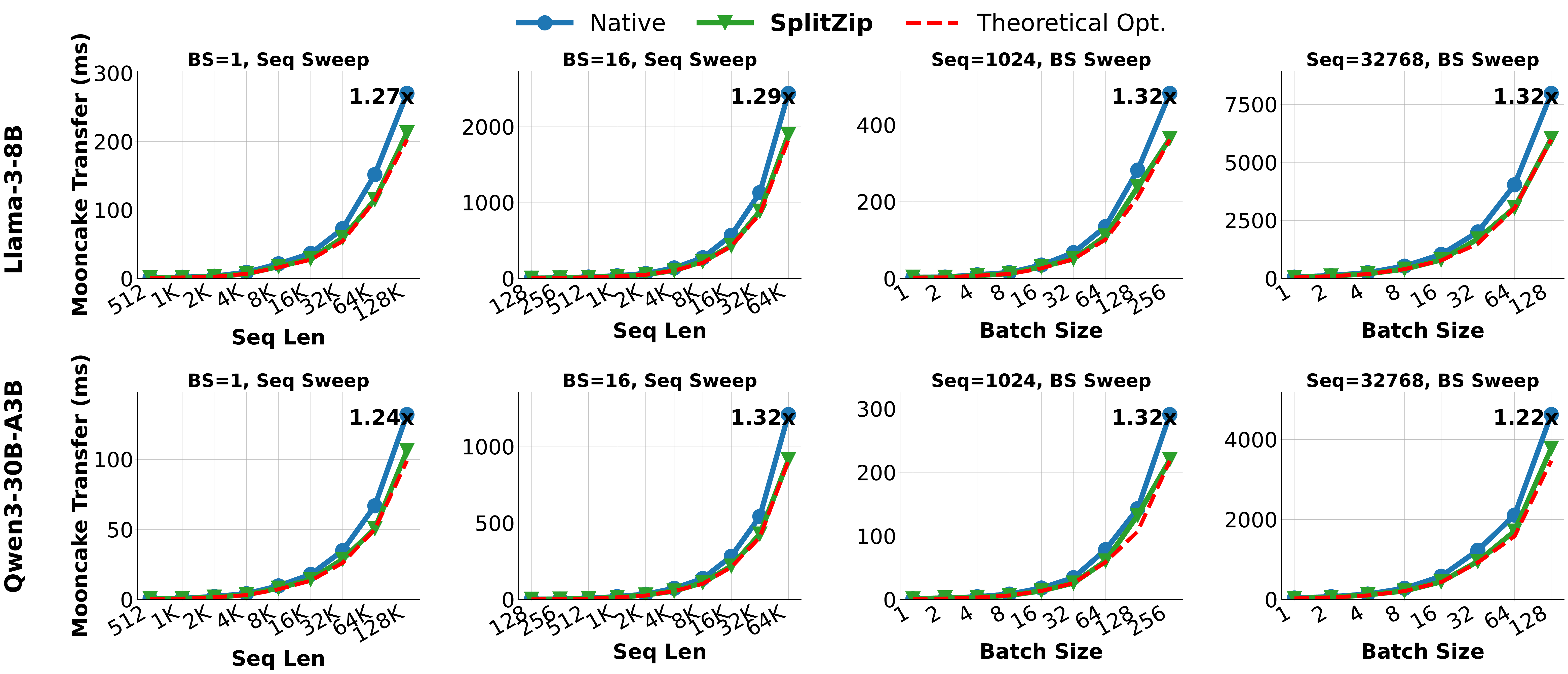}
        \caption{
        KV cache transfer time across sequence-length and batch-size sweeps using Mooncake\cite{qin2024mooncake}.
        }
        \label{fig:transfer_time_vs_seq_len}
        \vspace{-1.5em}
    \end{figure}
    
    \paragraph{KV Transfer time.}
    Figure~\ref{fig:transfer_time_vs_seq_len} reports the measured Mooncake transfer time for Llama-3-8B and Qwen3-30B-A3B across different batchSize and sequence lengths. 
    Theoretical Opt here represents the theoretically optimal time under the assumption of zero encoding/decoding overhead and the complete absence of any escape values.
    Across all eight panels, SplitZip consistently reduces Mooncake transfer time once the KV payload becomes large enough for bandwidth to dominate the end-to-end transfer cost.
    For short sequences, the benefit is smaller because fixed overheads and kernel launch costs account for a larger fraction of the total time.
    As sequence length increases, however, the transfer time becomes increasingly bandwidth-bound, and the reduction in transferred bytes directly translates into lower latency.
    At the largest point in each sweep, SplitZip achieves $1.27\times$--$1.32\times$ speedup on Llama-3-8B and $1.22\times$--$1.32\times$ speedup on Qwen3-30B-A3B.
    These results show that SplitZip provides consistent transfer-time reduction across model architectures and workload scales.

    Mooncake is the default KV-transfer backend for PD disaggregated serving in SGLang\cite{sglang2025}, a widely used LLM serving framework, and is therefore representative of practical deployments.
    We also integrate SplitZip into SGLang and evaluate end-to-end performance across sequence-length sweeps.
    
    As shown in Figure~\ref{fig:sglang_speedup}, SplitZip provides consistent improvements once the workload becomes transfer-dominated at longer sequence lengths, while slight slowdowns can occur in small-payload regimes due to fixed overheads.
    For batchSize=1 with sequence length ranging from 512 to 128K, SplitZip achieves TTFT speedups up to $1.303\times$ and throughput improvements up to $1.233\times$.
    For batchSize=16 with sequence length ranging from 128 to 64K, the TTFT ranges up to $1.274\times$, while requests throughput improvements up to $1.206\times$.  
    
    \begin{figure}[t]
        \centering
        \includegraphics[width=\textwidth]{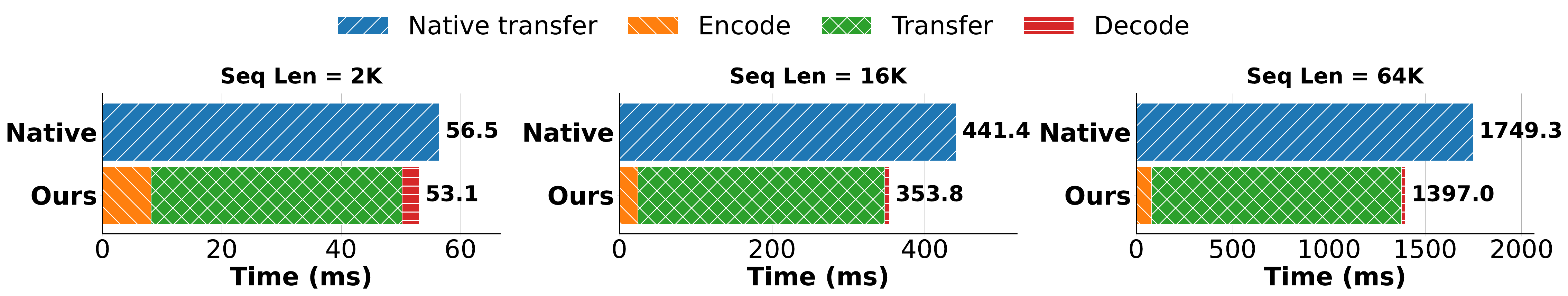}
        \caption{
        Transmission-time breakdown on Qwen3-32B.
        Native transfer sends the raw BF16 KV cache, while SplitZip consists of encoding, compressed transfer, and decoding under RoCE $4\times200$G.
        }
        \label{fig:qwen32_transmission_breakdown}
        \vspace{-1.5em}
    \end{figure}
    
    \paragraph{Transmission-time breakdown.}
    Figure~\ref{fig:qwen32_transmission_breakdown} further breaks down the Qwen3-32B transfer time with batchSize equal to 16.
    This breakdown uses additive accounting for encode, compressed transfer, and decode time.
    At sequence length $2$K, SplitZip slightly reduces transfer time from $56.5$~ms to $53.1$~ms.
    However, the advantage increases with sequence length.
    At $16$K, SplitZip reduces transfer time from $441.4$~ms to $353.8$~ms.
    At $64$K, it reduces transfer time from $1749.3$~ms to $1397.0$~ms.
    In this long-context regime, compressed transfer accounts for $92.9\%$ of the SplitZip time, while encoding and decoding account for only $5.7\%$ and $1.4\%$, respectively.
    This confirms that SplitZip keeps codec overhead small enough for long-context KV cache transfer to remain primarily communication-bound.


    
    

\subsection{Ablation Study}
    \subsubsection{Two-Pass Encoding}
\label{sec:two_pass}

SplitZip uses a two-pass encoding scheme to compress rare exponent escapes while avoiding reserving large amounts of GPU memory.
Although a one-pass encoder could eliminate the second scan, it must determine escape-buffer offsets before the total number of escapes is known. This entails either worst-case buffer preallocation or dynamic allocation using atomic operations, both of which introduce substantial overhead.

As shown in Table~\ref{tab:two_pass_memory}, worst-case provisioning requires $1.5\times$ the raw KV size, reaching $48$~GiB for Qwen3-32B at 131K context versus only $125.28$~MiB of actual escapes, overprovisioning by $392\times$.
This substantially constrains GPU memory capacity and concurrency. The alternative approach of, allocating escapes dynamically avoids worst-case reservation but reduces encoding throughput from $613.3$ to $491.2$~GB/s due to contention.
Therefore, we adopt the two-pass encoding scheme, facilitating exact buffer allocation and higher throughput than the dynamic allocation alternative.

\vspace{-1em}
\begin{table}[h]
\centering
\footnotesize
\caption{Escape-buffer provisioning for the two-pass encoder, which allocates from exact counts, versus a one-pass encoder. Worst-case one-pass provisioning must assume every KV value escapes.}
\vspace{0.5em}
\label{tab:two_pass_memory}
\setlength{\tabcolsep}{5pt}
\renewcommand{\arraystretch}{1.08}
\begin{tabular}{lcccc}
\toprule
Workload & Raw KV & Exact buffers & Worst-case one-pass & Over-provisioning \\
\midrule
Qwen3-32B, 8K context   & 2\,GiB  & 7.83\,MiB   & 3\,GiB   & $392\times$ \\
Qwen3-32B, 32K context  & 8\,GiB  & 31.32\,MiB  & 12\,GiB  & $392\times$ \\
Qwen3-32B, 131K context & 32\,GiB & 125.28\,MiB & 48\,GiB  & $392\times$ \\
Llama-3-8B, 8K context  & 1\,GiB  & 5.83\,MiB   & 1.5\,GiB & $263\times$ \\
\bottomrule
\end{tabular}
\end{table}

    \subsubsection{Ablation on Top-$k$ Exponent Coding}
\label{sec:ablation_topk}

Table~\ref{tab:ablation_topk} compares Top-8 3-bit coding with Top-16 4-bit coding.
Top-8 reduces the dense code width, but the much larger escape stream overwhelms this saving and lowers the measured compression ratio from $1.324\times$ to $1.241\times$.
Encoding throughput is similar for the two variants, while Top-16 is $3.07\times$ faster in decoding.

\begin{wraptable}{r}{0.33\linewidth}
    \vspace{-1.2em}
    \centering
    \caption{Top-$k$ exponent coding.}
    \label{tab:ablation_topk}
    \footnotesize
    \setlength{\tabcolsep}{4pt}
    \renewcommand{\arraystretch}{1.08}
    \begin{tabular}{lcc}
        \toprule
        \textbf{Metric} & \textbf{Top-8} & \textbf{Top-16} \\
        \midrule
        Code width & 3-bit & 4-bit \\
        Coverage & 92.11\% & 99.84\% \\
        Ratio & $1.241\times$ & $1.324\times$ \\
        Encode(GB/s) & 440.1 & 613.3 \\
        Decode(GB/s) & 710.5 & 2181.8 \\
        Escape rate & 7.89\% & 0.16\% \\
        \bottomrule
    \end{tabular}
    \vspace{-3em}
\end{wraptable}

The slowdown of Top-8 comes from both higher escape overhead and less GPU-friendly packing.
Reducing the codebook from 16 to 8 entries lowers coverage from 99.84\% to 92.11\%, increasing the escape rate from 0.16\% to 7.89\%.
This nearly $50\times$ higher escape rate increases memory-intensive escape collection and sparse overwrite costs.
The additional memory accesses and irregular correction operations further limit GPU parallelism and decoding throughput.
In addition, unlike 4-bit codes which can be byte-packed, 3-bit codes do not align naturally with byte-oriented GPU operations.


    \subsubsection{Ablation on Calibration Dataset}
\label{sec:ablation_calibration}

\begin{wraptable}{r}{0.53\linewidth}
    \vspace{-1.2em}
    \centering
    \caption{Cross-dataset calibration coverage. Dataset A is WikiText-2 train.}
    \label{tab:ablation_calibration}
    \scriptsize
    \setlength{\tabcolsep}{3pt}
    \renewcommand{\arraystretch}{1.02}
    \begin{tabular}{l|cc|cc|cc}
        \toprule
        \multirow{2}{*}{Eval set} 
        & \multicolumn{2}{c|}{Qwen3-32B}
        & \multicolumn{2}{c|}{Qwen3-Coder-30B}
        & \multicolumn{2}{c}{Llama-3-8B} \\
        \cmidrule(lr){2-3} \cmidrule(lr){4-5} \cmidrule(lr){6-7}
        & A$\rightarrow$B & B$\rightarrow$B 
        & A$\rightarrow$B & B$\rightarrow$B
        & A$\rightarrow$B & B$\rightarrow$B \\
        \midrule
        WikiText-2 & 99.83\% & 99.83\% & 99.48\% & 99.48\% & 99.85\% & 99.85\% \\
        HumanEval  & 99.81\% & 99.83\% & 99.44\% & 99.49\% & 99.84\% & 99.85\% \\
        GSM8K      & 99.79\% & 99.80\% & 99.41\% & 99.45\% & 99.78\% & 99.79\% \\
        MMLU       & 99.78\% & 99.79\% & 99.38\% & 99.33\% & 99.77\% & 99.80\% \\
        PTB        & 99.76\% & 99.76\% & 99.34\% & 99.34\% & 99.73\% & 99.74\% \\
        \bottomrule
    \end{tabular}
    \vspace{-1.7em}
\end{wraptable}

SplitZip uses a calibrated top-16 exponent codebook to avoid online histogram construction.
To test whether this codebook is dataset-specific, we calibrate on WikiText-2 train and evaluate coverage on datasets from different domains.
Table~\ref{tab:ablation_calibration} compares cross-dataset calibration, denoted A$\rightarrow$B, with oracle calibration on each target dataset, denoted B$\rightarrow$B.

The calibrated exponent set generalizes well.
Coverage remains higher than 99\% across all evaluated datasets, and almost matches the oracle result on WikiText-2 test, HumanEval~\cite{chen2021humaneval}, GSM8K~\cite{cobbe2021gsm8k}, MMLU~\cite{hendrycks2021mmlu}, and PTB~\cite{zhou2020ptb} in this run.
This shows that high-frequency KV cache exponents are stable across domains, enabling SplitZip to use a fixed offline codebook.

    \subsubsection{Ablation on Calibration Granularity}
\label{sec:ablation_granularity}

\begin{wraptable}[6]{r}{0.40\linewidth}
    \vspace{-1.6em}
    \centering
    \caption{Calibration granularity ablation.}
    \label{tab:ablation_granularity}
    \scriptsize
    \setlength{\tabcolsep}{5pt}
    \renewcommand{\arraystretch}{1.02}
    \begin{tabular}{lccc}
        \toprule
        \textbf{Metric} & \textbf{Tensor} & \textbf{Token} & \textbf{Channel} \\
        \midrule
        Coverage & 99.84\% & 99.90\% & 99.90\% \\
        Compression & $1.324\times$ & $1.324\times$ & $1.327\times$ \\
        Encode(GB/s) & 613.3 & 0.082 & 0.020 \\
        Decode(GB/s) & 2181.8 & 0.173 & 0.060 \\
        \bottomrule
    \end{tabular}
    \vspace{-1.0em}
\end{wraptable}

Some prior compression designs use finer-grained codebooks to improve locality, compression ratio, or parallelism.
We evaluate this idea on per-tensor, per-token and per-channel codebooks.

Table~\ref{tab:ablation_granularity} shows that finer-grained calibration increases top-16 coverage, reaching 99.90\% with per-token and per-channel codebooks.
However, throughput drops by several orders of magnitude: per-token and per-channel calibration both fall from hundreds of GB/s to sub-GB/s performance.
This is because fine-grained calibration requires many small codebooks, introducing extra storage, irregular lookup patterns, and more complex GPU execution.
    \subsubsection{Ablation on Escape-Position Metadata}
\label{sec:ablation_escape_position}

\begin{wraptable}{r}{0.4\linewidth}
    \vspace{-1.4em}
    \centering
    \caption{Escape-position ablation.}

    \label{tab:ablation_escape_position}
    \scriptsize
    \setlength{\tabcolsep}{6pt}
    \renewcommand{\arraystretch}{1.08}
    \begin{tabular}{lcc}
        \toprule
        \textbf{Metric} & \textbf{Top-16 + Pos.} & \textbf{Top-15 + Sent.} \\
        \midrule
        Coverage & 99.84\% & 99.73\% \\
        Escape rate & 0.16\% & 0.27\% \\
        Ratio & $1.324\times$ & $1.331\times$ \\
        Encode(GB/s) & 613.3  & 396.0  \\
        Decode(GB/s) & 2181.8  & 620.8  \\
        \bottomrule
    \end{tabular}
    \vspace{-1.4em}
\end{wraptable}

We also study whether SplitZip should explicitly store escape positions or reserve one code as an escape sentinel.
The sentinel design uses only 15 frequent exponent values and reserves the remaining 4-bit code to mark escaped elements.
This avoids storing explicit chunk-local positions and therefore slightly improves the compression ratio, from $1.324\times$ to $1.331\times$.
However, this metadata saving comes at a large decoding cost.

As shown in Table~\ref{tab:ablation_escape_position}, the Top-15 sentinel design reduces decode throughput from 2181.8~GB/s to 620.8~GB/s, a $3.5\times$ slowdown.
The reason is that the decoder can no longer treat all 4-bit codes uniformly.
It must inspect the dense code stream to identify sentinel values and then merge escaped exponents back into the output, introducing irregular control flow and memory access.
In contrast, the Top-16 design decodes every element through the same dense lookup path and applies rare escape corrections through a separate sparse overwrite.
This requires explicit position, making the metadata overhead a little higher.
SplitZip prioritizes throughput, where the minimal cost of additional metadata is amortized by the higher throughput, making Top-16 coding the natural choice.

    \subsubsection{Ablation on Pre-Calibration}
\label{sec:ablation_precalibration}

\begin{wraptable}[8]{r}{0.40\linewidth}
    \vspace{-1.2em}
    \centering
    \caption{Pre-calibration ablation.}
    \label{tab:ablation_precalibration}
    \scriptsize
    \setlength{\tabcolsep}{8pt}
    \renewcommand{\arraystretch}{1.08}
    \begin{tabular}{lcc}
        \toprule
        \textbf{Metric} & \textbf{Pre-calib.} & \textbf{Dynamic} \\
        \midrule
        Ratio & $1.324\times$ & $1.324\times$ \\
        Escape rate & 0.16\% & 0.16\% \\
        Encode(GB/s) & 613.3 & 80.7 \\
        Decode(GB/s) & 2181.8 & 2197.3 \\
        \bottomrule
    \end{tabular}
    \vspace{-1.5em}
\end{wraptable}

SplitZip's offline pre-calibrated codebook is compared with a dynamic Top-16 variant that rebuilds the exponent codebook for each input in Table~\ref{tab:ablation_precalibration}. Dynamic calibration achieves the same compression ratio and escape rate as the pre-calibrated design, and decode throughput is almost unchanged.
However, it substantially slows down online compression: encode throughput drops from 613.3~GB/s to 80.7~GB/s, $7.6\times$ slower, because the dynamic variant adds an online histogram and top-$k$ selection pass.

    \subsubsection{Ablation on Quad64 Vectorized Encoding}
\label{sec:ablation_quad64}

    \label{sec:quad64}
    
    \begin{wraptable}[7]{r}{0.42\linewidth}
        \vspace{-1.2em}
        \centering
        \caption{Quad64 vectorization ablation.}
        \label{tab:ablation_quad64}
        \scriptsize
        \setlength{\tabcolsep}{7pt}
        \renewcommand{\arraystretch}{1.08}
        \begin{tabular}{lcc}
            \toprule
            \textbf{Variant} & \textbf{Encode} & \textbf{Speedup} \\
            & \textbf{GB/s} & \\
            \midrule
            Scalar pair encode & 416.2 & $1.00\times$ \\
            Quad64 encode & 613.3 & $1.47\times$ \\
            \bottomrule
        \end{tabular}
        \vspace{-1.5em}
    \end{wraptable}

    As shown in Table~\ref{tab:ablation_quad64}, Quad64 improves encode throughput from 416.2~GB/s to 613.3~GB/s. This corresponds to a $1.47\times$ speedup over the scalar approach. 
    Thus, the Quad64-vectorization based optimization is critical in enabling  SplitZip to reach its final high-throughput online compression performance.
    \subsubsection{Ablation on Layer-Wise Coverage}
\label{sec:ablation_layer_coverage}


\begin{wrapfigure}{r}{0.43\linewidth}
    \vspace{-1.3em}
    \centering
    \caption{Layer-wise coverage under a fixed shared Top-16 codebook on Qwen3-32B.}
    \label{tab:ablation_layer_coverage}
    \includegraphics[width=\linewidth]{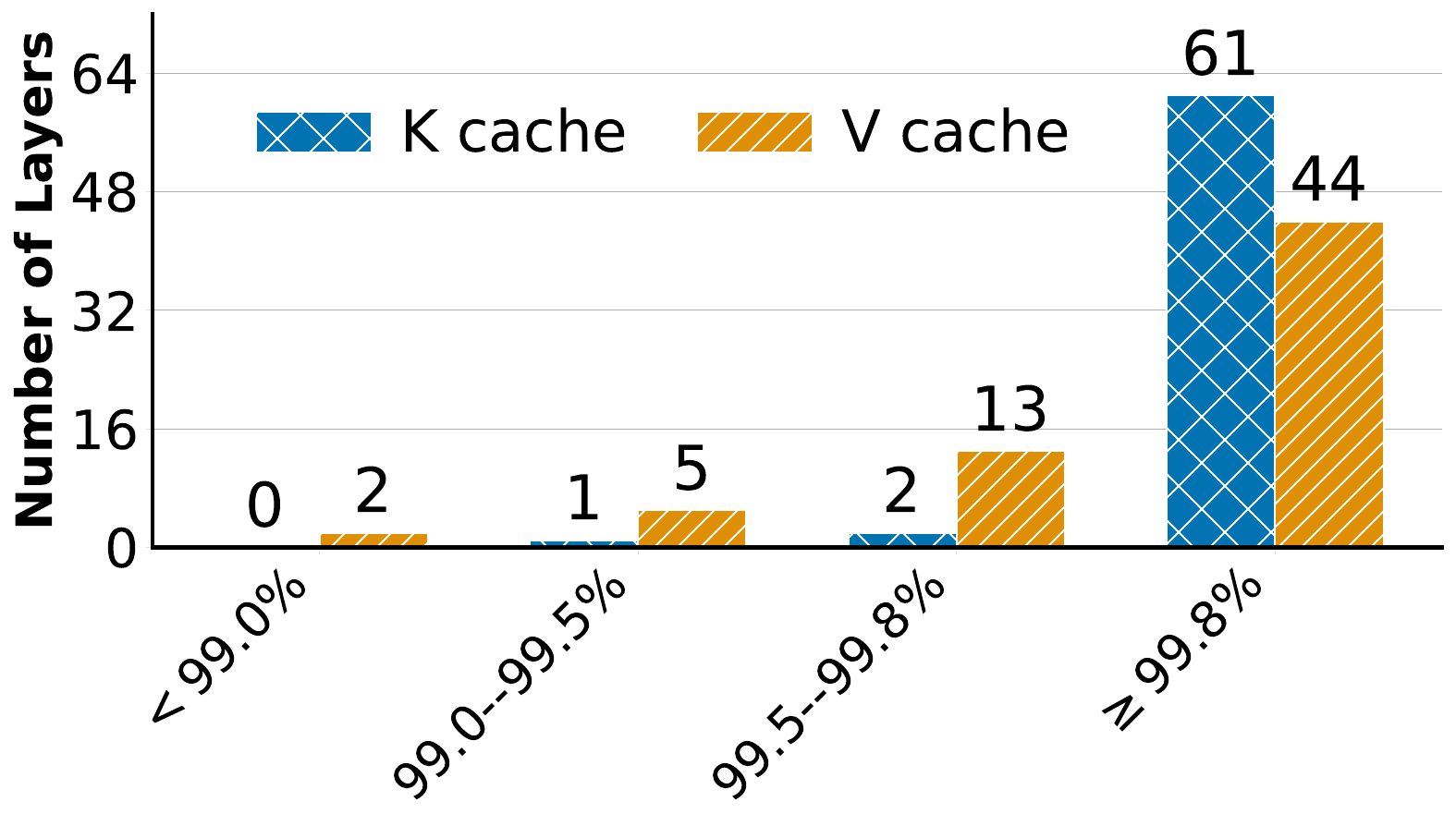}
    \vspace{-3em}
\end{wrapfigure}

We further test whether a small number of layers exhibit heavier-tailed exponent distributions that would increase the escape rate.
For Qwen3-32B, we profile BF16 KV caches from all 64 layers, select one shared Top-16 codebook from the aggregate K-cache distribution and one shared Top-16 codebook from the aggregate V-cache distribution, and then apply each fixed codebook back to every layer.
Figure~\ref{tab:ablation_layer_coverage} reports the resulting layer-wise coverage histogram.
It directly measures whether the deployed fixed codebook produces low-coverage outlier layers.

The fixed K-cache codebook is stable across layers: all 64 layers remain above 99.0\% coverage, and 61 layers remain above 99.8\%.
The V-cache codebook shows a modest lower-coverage tail, with two early layers below 99.0\% and a worst-layer coverage of 98.77\%.
However, even this worst case corresponds to a 1.23\% escape rate, and the median V layer still reaches 99.88\% coverage. While per-layer codebooks would eliminate this tail, the cost in codebook management and per-layer storage might offset the throughput advantage of the global codebook design.

\section{Conclusion}
\label{sec:conclusion}

This paper presents SplitZip, a GPU-friendly lossless compression scheme for KV cache transfer in PD disaggregated LLM serving.
SplitZip encodes frequent exponent values with fixed-length codes, and corrects rare values through an explicit escape stream.
Our experiments show that SplitZip achieves higher compression and decompression throughput than prior lossless codecs, and translates this throughput into end-to-end KV-transfer speedups for long-context workloads.

SplitZip is most effective when KV cache communication is a major bottleneck; for short contexts or compute-bound regimes, codec overhead may outweigh the reduced transfer volume. Its compression ratio is also bounded by the redundancy in floating-point exponent values, making it complementary to more aggressive lossy or model-aware compression methods. In particular, SplitZip can be applied on top of an already-quantized KV cache (e.g., FP8) to recover additional bandwidth savings without further accuracy risk, as demonstrated by our E5M2 results. 

SplitZip reduces the cost, latency, and energy footprint of long-context LLM serving without changing model outputs. As with any efficiency improvement, this may also increase total deployment volume, so it should be used together with appropriate governance.

\begin{ack}
This material is based upon work supported by the National Science Foundation under CAREER Award No. CCF-2340799. Yipin Guo was supported in part by an Arthur J. Schmitt Presidential Leadership Fellowship at the University of Notre Dame. We also acknowledge DFloat11 and Huff-LLM for inspiring this work.
\end{ack}

{\small
\bibliographystyle{plainnat}
\bibliography{references}
}

\newpage
\appendix

\section{Why Losslessness Matters: BF16 versus FP8 E4M3 KV Caches}
\label{sec:lossy_fp8}

\vspace{-2em}
\begin{table}[h]
\centering
\small
\caption{Task accuracy with BF16 versus FP8 E4M3 prefill KV caches. \textsc{Qwen3-32B-W4} denotes Qwen3-32B with 4-bit
quantized weights.}
\label{tab:fp8_accuracy}
\setlength{\tabcolsep}{6pt}
\renewcommand{\arraystretch}{1.08}
\begin{tabular}{llcc}
\toprule
Benchmark & Model & BF16 KV & FP8 E4M3 KV \\
\midrule
LongBench-v2       & Qwen3-32B          & 43.61\% & 42.11\% \\
LongBench-v2       & Qwen3-32B-W4       & 40.60\% & 39.10\% \\
LongBench-v2       & Qwen3.6-35B-A3B    & 51.56\% & 46.88\% \\
LongBench-v2       & Qwen3-30B-A3B      & 32.81\% & 28.13\% \\
LiveCodeBench Hard & Qwen3-32B          & 50.00\% & 35.71\% \\
\bottomrule
\end{tabular}
\end{table}

A natural question is why a lossless codec is worth deploying at all, given that simply storing the KV cache in FP8 halves the transferred payload.
Table~\ref{tab:fp8_accuracy} evaluates the effect of quantizing the prefill KV
cache to FP8 E4M3 while leaving model execution otherwise unchanged. Accuracy
degrades in all five configurations. The loss is modest but consistent on
LongBench-v2 for the dense Qwen3-32B model, and substantially larger for the two mixture-of-expert models. The largest degradation appears on LiveCodeBench Hard, where accuracy drops by $14.3\%$ percentage points.

\section{Codebook Coverage under Broader Distribution Shift}
\label{sec:coverage_shift}

\splitzip replaces online histogramming with a codebook calibrated once, offline
(Section~\ref{sec:method_calibration}). The deployment risk this creates is
distribution shift: if production traffic drives the KV exponent distribution away
from the calibration corpus, coverage falls, the escape stream grows, and the
compression ratio degrades. Correctness is never at stake, since uncovered
exponents are stored exactly, but throughput and payload size are. Section~\ref{sec:ablation_calibration}
tests four evaluation corpora; this appendix extends the test to a substantially
wider set of input regimes, to cross-model codebook reuse, and to the cost of
simply recalibrating.


\begin{table}[h]
\centering
\small
\caption{Fixed Top-16 coverage on Qwen3-32B for a codebook calibrated only on
WikiText, evaluated across languages, domains, and degenerate input structures.}
\label{tab:coverage_shift}
\setlength{\tabcolsep}{7pt}
\renewcommand{\arraystretch}{1.08}
\begin{tabular}{llcc}
\toprule
Evaluation regime & Character & Fixed Top-16 coverage & Realized ratio \\
\midrule
English exposition  & in-domain (calibration corpus) & 99.856\% & $1.330\times$ \\
Python code         & domain shift                   & 99.839\% & $1.329\times$ \\
Math reasoning      & domain shift                   & 99.840\% & $1.329\times$ \\
Instruction/chat    & format shift                   & 99.847\% & $1.330\times$ \\
Chinese             & language shift                 & 99.754\% & $1.327\times$ \\
Arabic              & language shift                 & 99.780\% & $1.328\times$ \\
Long-document QA    & length shift                   & 99.839\% & $1.329\times$ \\
Repetitive input    & degenerate structure           & 99.833\% & $1.329\times$ \\
Unusual format      & degenerate structure           & 99.828\% & $1.329\times$ \\
\bottomrule
\end{tabular}
\end{table}

Table~\ref{tab:coverage_shift} evaluates a single WikiText-calibrated Top-16
codebook on nine input regimes spanning three kinds of shift: domain (code, math),
language (Chinese, Arabic), and structure (instruction formatting, long-document
QA, highly repetitive text, and deliberately unusual formatting). Coverage stays
above $99.75\%$ everywhere, and the full spread across all nine regimes is
$0.102$ percentage points, from $99.754\%$ to $99.856\%$. 

The language shifts produce the lowest coverage of the nine, which is the
expected direction, but the effect is small in absolute terms. We read this as
evidence that the frequent exponents of a KV cache are determined mainly by the
numerical scale structure of the trained model, namely the layer norms and
projection magnitudes that set the dynamic range of the activations, rather than
by the semantic content of the input. Ordinary production drift, in which a
deployment gradually serves a different mix of languages or domains than it was
calibrated on, is therefore unlikely to move the operating point materially.

\section{The Potential of Hiding Codec Overhead with Pipeline Overlap}

    In prefill--decode disaggregated serving, KV cache compression, network transfer, and decompression can be executed as a streaming pipeline. 
    This is important because the latency of a standalone codec kernel does not necessarily translate into additional end-to-end latency: if communication is the bottleneck, the encode and decode stages can be overlapped with data transfer.
    
    Let $S$ denote the raw KV cache size, $\rho$ the compression ratio, $G_{\mathrm{enc}}$ the compression throughput, $G_{\mathrm{dec}}$ the decompression throughput, and $B$ the physical communication bandwidth for transferring compressed bytes. 
    For one pipeline chunk, the three stage times are
    \[
        T_{\mathrm{enc}} = \frac{S}{G_{\mathrm{enc}}},
        \qquad
        T_{\mathrm{xfer}} = \frac{S}{\rho B},
        \qquad
        T_{\mathrm{dec}} = \frac{S}{G_{\mathrm{dec}}}.
    \]
    For a long stream of chunks, the steady-state pipeline time is dominated by the slowest stage:
    \[
        T_{\mathrm{pipe}}
        =
        \max\left(
            \frac{S}{G_{\mathrm{enc}}},
            \frac{S}{\rho B},
            \frac{S}{G_{\mathrm{dec}}}
        \right).
    \]
    
    The codec overhead is fully hidden when the transfer stage remains the bottleneck, i.e.
        $T_{\mathrm{xfer}} \ge T_{\mathrm{enc}}
        \quad \text{and} \quad
        T_{\mathrm{xfer}} \ge T_{\mathrm{dec}}$.
    Equivalently, the maximum physical communication bandwidth for fully hiding both encoding and decoding overhead is
    \[
        B_{\mathrm{hide}}
        =
        \frac{\min(G_{\mathrm{enc}}, G_{\mathrm{dec}})}{\rho}.
    \]
    Thus, a faster codec increases the range of communication bandwidths under which SplitZip behaves as a purely bandwidth-saving transformation with no exposed codec latency.
    
    In our implementation, SplitZip achieves a compression throughput of $G_{\mathrm{enc}} = 613.3~\mathrm{GB/s}$ and a decompression throughput of
    $G_{\mathrm{dec}} = 2181.8~\mathrm{GB/s}$. Since compression is the slower codec stage, the hiding threshold is determined by $G_{\mathrm{enc}}$:
    \[
        B_{\mathrm{hide}}
        =
        \frac{\min(613.3, 2181.8)}{1.324}
        =
        \frac{613.3}{1.324}
        \approx
        463.2~\mathrm{GB/s}.
    \]
    
    Under an idealized streaming pipeline with complete stage overlap, codec overhead could in principle be hidden when the physical communication bandwidth is below approximately $463.2$~GB/s. 
    This analysis represents the potential of pipeline overlap rather than an optimization implemented in our measured transfer pipeline.
    
    This threshold is well above the bandwidth of common inter-node communication technologies. For example, 400~Gb/s and 800~Gb/s Ethernet correspond to approximately $50~\mathrm{GB/s}$ and $100~\mathrm{GB/s}$, respectively, and PCIe~5.0~x16 provides roughly $64~\mathrm{GB/s}$ per direction. 
    Thus, under typical datacenter interconnects, SplitZip's encoding and decoding overheads can in principle be fully covered by communication time. 
    In this regime, SplitZip behaves as a bandwidth-saving transformation rather than adding exposed codec latency to the end-to-end prefill--decode transfer path.
    
    Only very high-bandwidth intra-node GPU fabrics, such as modern NVLink-class interconnects, approach or exceed this threshold. In those settings, communication may no longer be the sole bottleneck, and codec throughput can again become visible in the end-to-end critical path.

\section{SplitZip on FP8}
    \label{sec:fp8_result}

\begin{table}[t]
\centering
\small
\caption{FP8 codec throughput and compression ratio.}
\label{tab:fp8_throughput}
        \begin{tabular}{@{}lccc@{}}
            \toprule
            \textbf{Metric}
            & \textbf{E4M3}
            & \textbf{E5M2}
            & \textbf{E5M2} \\
            & \textbf{Top-8}
            & \textbf{Top-8}
            & \textbf{Top-16} \\
            \midrule
            Coverage
            & 92.17\%
            & 92.28\%
            & 99.84\% \\
            Ratio vs. FP8
            & $0.933\times$
            & $1.049\times$
            & $1.136\times$ \\
            Ratio vs. BF16
            & $1.866\times$
            & $2.097\times$
            & $2.273\times$ \\
            Encode(GB/s)
            & 219.6$\pm$1.5
            & 221.6$\pm$14.6
            & 249.7$\pm$0.8 \\
            Decode(GB/s)
            & 366.9$\pm$2.5
            & 340.8$\pm$0.9
            & 564.9$\pm$6.5 \\
            Escape rate
            & 7.83\%
            & 7.72\%
            & 0.16\% \\
            \bottomrule
        \end{tabular}
\end{table}

    SplitZip can also be applied to FP8 KV caches by adapting the exponent code width to the FP8 format. 
    We consider both E5M2 and E4M3. 
    For E5M2, the 5-bit exponent can be compressed with either a 4-bit top-16 code or a 3-bit top-8 code. For E4M3, a 4-bit code would not reduce the exponent size, so SplitZip uses a 3-bit top-8 code.
    FP8 offers less redundancy than BF16 because the original representation is already smaller, but the same fixed-code-plus-escape principle still applies.
    
    Table~\ref{tab:fp8_throughput} reports the FP8 results of SplitZip.
    Since FP8 already halves the KV cache size compared with BF16, the remaining lossless compression opportunity is smaller.
    Nevertheless, SplitZip still provides additional compression over native E5M2 FP8.
    E4M3 Top-8 expands the payload because its exponent field is already compact and the escape metadata is too large for this distribution.
    E5M2 offers more redundancy, and the Top-16 variant achieves both the highest compression ratio and the lowest escape rate.
    
    The lower escape rate of E5M2 Top-16 also improves throughput: it reaches 249.7~GB/s encoding and 564.9~GB/s decoding, compared with 221.6~GB/s and 340.8~GB/s for E5M2 Top-8.
    This mirrors the BF16 trend that reducing escapes is more important than shrinking the dense code width when sparse correction dominates the tail cost.
    We therefore recommend Top-16 as the default FP8 setting for the measured Qwen3-32B workload.

\section{Ablation on Chunk Size}
\label{sec:ablation_chunksize}


\begin{wrapfigure}{r}{0.43\linewidth}
    \vspace{-1.3em}
    \centering
    \includegraphics[width=\linewidth]{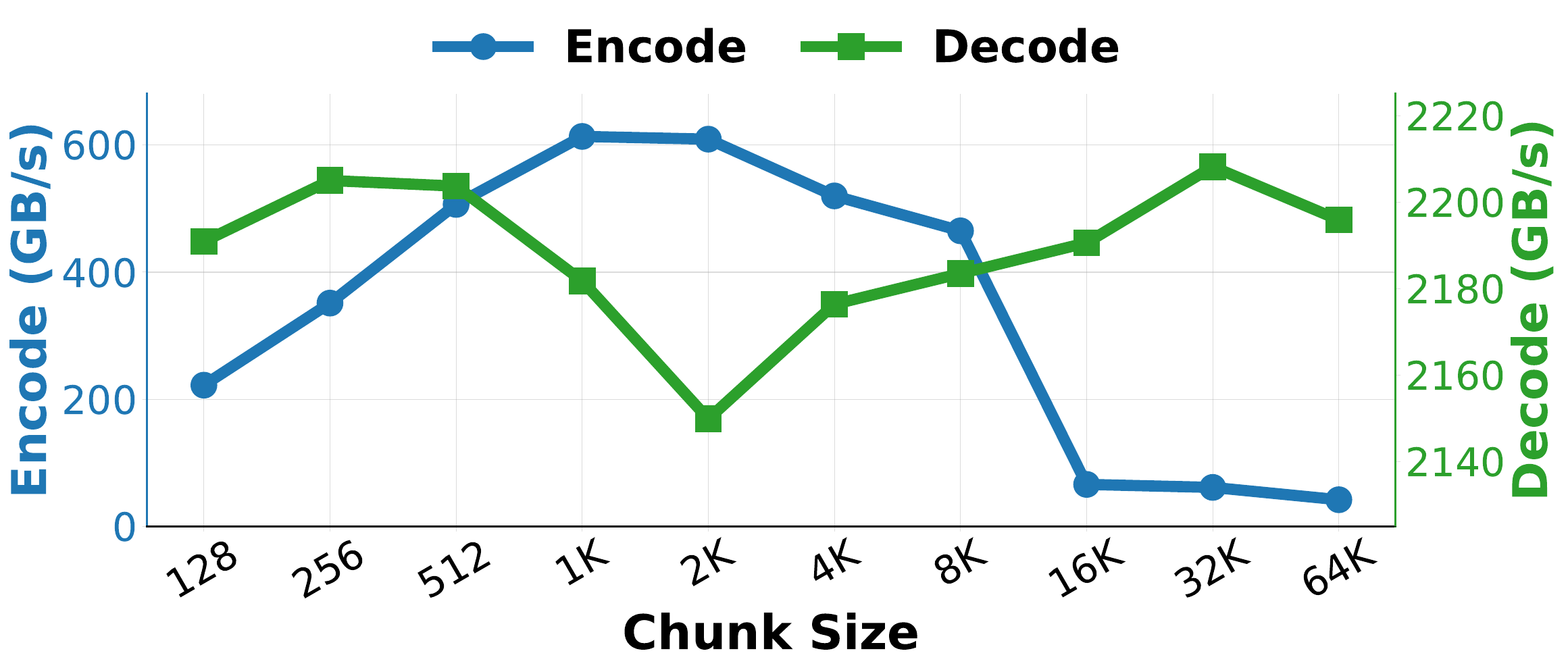}
        \caption{Escape value capture chunk-size ablation.}    \label{fig:ablation_chunksize}
    \vspace{-1em}
\end{wrapfigure}

We evaluate the sensitivity of SplitZip to the escape-collection chunk size.
The chunk size determines the local range used to record escape positions.
When the chunk size is no larger than 256, each escape position can be represented with an 8-bit integer
instead of a 16-bit integer, reducing the metadata cost of the sparse escape stream.
However, this metadata saving comes at a substantial throughput cost.
As shown in Figure~\ref{fig:ablation_chunksize}, using chunk size 256 reduces encode throughput from
613.3~GB/s to 351.3~GB/s.
The reason is that smaller chunks increase the number of chunks that must be processed, which introduces
more per-chunk bookkeeping, more metadata operations, and less efficient escape collection.
Although the uint8 escape-position representation slightly reduces compressed size, the escape rate is already
very low under Top-16 coding, so the absolute metadata saving is small.

Larger chunks do not always improve performance.
While chunk size 2048 achieves similar throughput to 1024, further increasing the chunk size gradually reduces encode throughput, and very large chunks such as 16K or above cause a sharp slowdown.
Overall, chunk size 1024 provides the best balance between compact metadata, high encode throughput,
and high decode throughput.

In addition, the chunked representation itself is important for decoding efficiency.
Recording escape positions as chunk-local offsets substantially improves decode throughput; if escape positions are instead stored as 32-bit absolute indices, decode throughput drops to \textit{1421.7~GB/s}, while the representation also risks integer overflow for very large KV tensors.
Therefore, considering both metadata size and codec throughput, SplitZip uses chunk size 1024 as the default setting.

\section{Codec Throughput on More GPUs}
\label{sec:more_gpu}

\begin{table*}[t]
\centering
\caption{Codec throughput across different NVIDIA GPUs. The compression ratio is reported using the H200 measurement as the reference. Each throughput entry is reported as encoding/decoding throughput in GB/s.}
\label{tab:codec_gpu_comparison}
\setlength{\tabcolsep}{2pt}
\resizebox{0.9\linewidth}{!}{
\begin{tabular}{lccccc}
\toprule
Method & Ratio & H200 & A100 & H100 & B200 \\
\midrule
nvCOMP LZ4~\cite{nvidia2026nvcomp}
& 1.019 & $13.4/137.1$ & $6.2/31.7$ & N/A & $25.9/77.6$ \\

nvCOMP Cascaded~\cite{nvidia2026nvcomp}
& 1.000 & $111.8/155.2$ & $52.1/32.4$ & $101.6/176.3$ & $108.7/109.9$ \\

nvCOMP Bitcomp~\cite{nvidia2026nvcomp}
& 1.056 & $341.5/147.7$ & $116.4/33.9$ & $290.8/180.4$ & $258.3/111.3$ \\

ZipNN~\cite{zipnn}
& 1.515 & $1.2/1.7$ & 1.2/1.7 & 1.2/1.7 & 1.2/1.7 \\

DFloat11~\cite{dfloat11}
& 1.423 & $4.0e\!-\!03/468.2$ & $4.28e\!-\!03/224.3$ & $6.6e\!-\!03/496.3$ & $4.108e\!-\!03/515.7$ \\

Falcon~\cite{Li2025falcon}
& 1.428 & $8.9/14.4$ & $4.3/6.0$ & $8.5/13.3$ & $8.9/15.2$ \\

ZipServ - CPP~\cite{fan2026zipserv}
& 1.236 & $0.1/499.5$ & $0.073/1.9$ & $0.0796/1.5$ & $0.0449/10.8$ \\

ZipServ - Kernel~\cite{fan2026zipserv}
& 1.236 & N/A$/1260.9$ & N/A$/791.1$ & N/A$/1156.5$ & N/A$/1493.0$ \\

\textbf{SplitZip}
& \textbf{1.324}
& \textbf{$613.3/2181.8$}
& \textbf{$352.5/842.6$}
& \textbf{$579.8/1486.3$}
& \textbf{$637.9/2607.4$} \\
\bottomrule
\end{tabular}}
\vspace{0.3em}
\begin{flushleft}
\footnotesize
Each entry is reported as encode/decode throughput in GB/s. N/A indicates that the corresponding encode or decode path is unavailable.
\end{flushleft}
\end{table*}

Table~\ref{tab:codec_gpu_comparison} shows that SplitZip consistently maintains the highest encoding throughput across all evaluated GPUs. Compared with the fastest available encoder, nvCOMP Bitcomp, SplitZip improves encoding throughput by $1.8\times$ on H200, $3.0\times$ on A100, $2.0\times$ on H100, and $2.5\times$ on B200. This confirms that the fixed-length exponent coding design scales robustly across GPU generations and avoids the CPU-side or variable-length encoding bottlenecks observed in prior lossless compressors.

For decoding, SplitZip also achieves the best throughput in all settings, but the relative gain depends more strongly on the underlying hardware. This is because the decode path is largely memory-movement bound: once the implementation approaches the hardware communication/memory-bandwidth limit, further algorithmic improvements provide diminishing returns. As a result, on A100 and H100, where ZipServ's kernel decoder already reaches a high fraction of the attainable bandwidth, SplitZip provides a relatively modest decode advantage. In contrast, on newer GPUs with higher bandwidth and stronger memory subsystems, SplitZip continues to scale: on B200, it reaches $2607.4$~GB/s decoding throughput, achieving a much larger advantage over ZipServ Kernel's $1493.0$~GB/s. These results suggest that SplitZip's lightweight and highly parallel decoding design is especially beneficial on modern high-bandwidth GPUs, where the codec can better exploit the available hardware throughput.

\section*{NeurIPS Paper Checklist}

\begin{enumerate}

\item {\bf Claims}
    \item[] Question: Do the main claims made in the abstract and introduction accurately reflect the paper's contributions and scope?
    \item[] Answer: \answerYes{}.
    \item[] Justification: The abstract and introduction distinguish controlled microbenchmarks, real-KV Mooncake measurements, lossless BF16 reconstruction, and the limitations of each setting.

\item {\bf Limitations}
    \item[] Question: Does the paper discuss the limitations of the work performed by the authors?
    \item[] Answer: \answerYes{}.
    \item[] Justification: In Section~\ref{sec:conclusion}.

\item {\bf Theory assumptions and proofs}
    \item[] Question: For each theoretical result, does the paper provide the full set of assumptions and a complete proof?
    \item[] Answer: \answerNA{}.
    \item[] Justification: The paper is an empirical systems paper and does not claim new theoretical results beyond the direct bitwise reconstruction property of the lossless codec.

\item {\bf Experimental result reproducibility}
    \item[] Question: Does the paper fully disclose all information needed to reproduce the main experimental results to the extent that it affects the main claims?
    \item[] Answer: \answerYes{}.
    \item[] Justification: The experiments section specifies hardware, models, settings, baselines.

\item {\bf Open access to data and code}
    \item[] Question: Does the paper provide open access to the data and code, with sufficient instructions to faithfully reproduce the main experimental results?
    \item[] Answer: \answerYes{}.
    \item[] Justification: The repository includes the SplitZip implementation and experiment scripts used for profiling, baseline comparisons. Please refer to Supplementary Material. It will be open-sourced after the article is accepted.

\item {\bf Experimental setting/details}
    \item[] Question: Does the paper specify all training and test details necessary to understand the results?
    \item[] Answer: \answerYes{}.
    \item[] Justification: No model training is performed. The test setup, evaluated models, context lengths, transport stack, precision modes, and measured quantities are described in Section~\ref{sec:experiments}.

\item {\bf Experiment statistical significance}
    \item[] Question: Does the paper report error bars suitably and correctly defined or other appropriate information about statistical significance?
    \item[] Answer: \answerYes{}.
    \item[] Justification: The encoding and decoding throughput of the method may be influenced by randomness; therefore, multiple measurements were conducted, and the measurement fluctuations are reported.

\item {\bf Experiments compute resources}
    \item[] Question: For each experiment, does the paper provide sufficient information on compute resources needed to reproduce the experiments?
    \item[] Answer: \answerYes{}.
    \item[] Justification: In Section \ref{sec:setup}.

\item {\bf Code of ethics}
    \item[] Question: Does the research conducted in the paper conform, in every respect, with the NeurIPS Code of Ethics?
    \item[] Answer: \answerYes{}.
    \item[] Justification: The work studies compression and systems performance for existing LLM serving workloads and does not involve human subjects, private data collection, or model release.

\item {\bf Broader impacts}
    \item[] Question: Does the paper discuss both potential positive societal impacts and negative societal impacts of the work performed?
    \item[] Answer: \answerYes{}.
    \item[] Justification: In Section~\ref{sec:conclusion}.

\item {\bf Safeguards}
    \item[] Question: Does the paper describe safeguards for responsible release of data or models that have a high risk for misuse?
    \item[] Answer: \answerNA{}.
    \item[] Justification: The paper does not release a new model, dataset, or user-facing generative system; it releases a compression method for KV tensors.

\item {\bf Licenses for existing assets}
    \item[] Question: Are creators or original owners of assets used in the paper properly credited and are licenses and terms of use explicitly mentioned and respected?
    \item[] Answer: \answerYes{}.
    \item[] Justification: The draft cites related baseline and datasets. All authors are listed in shared order..

\item {\bf New assets}
    \item[] Question: Are new assets introduced in the paper well documented and is the documentation provided alongside the assets?
    \item[] Answer: \answerYes{}.
    \item[] Justification: The paper introduces SplitZip and the accompanying implementation. The potentially released repository will document how to use the codec and include scripts for profiling codec throughput, evaluating compression behavior, and reproducing end-to-end KV cache transfer benchmarks.

\item {\bf Crowdsourcing and research with human subjects}
    \item[] Question: For crowdsourcing experiments and research with human subjects, does the paper include the full text of instructions given to participants and compensation details?
    \item[] Answer: \answerNA{}.
    \item[] Justification: This work does not involve crowdsourcing, user studies, or experiments with human subjects.

\item {\bf Institutional review board approvals or equivalent for research with human subjects}
    \item[] Question: Does the paper describe potential risks to study participants and whether IRB approvals were obtained?
    \item[] Answer: \answerNA{}.
    \item[] Justification: This work does not involve human-subject research, and therefore IRB approval or equivalent review is not applicable.

\item {\bf Declaration of LLM usage}
    \item[] Question: Does the paper describe the usage of LLMs if it is an important, original, or non-standard component of the core methods in this research?
    \item[] Answer: \answerNA{}.
    \item[] Justification: LLMs are used as standard serving workloads for evaluating KV cache transfer and compression performance. They are not used as an original, non-standard, or decision-making component of the proposed compression method.

\end{enumerate}

\end{document}